\documentclass[]{spie}  
\usepackage[]{graphicx}
\usepackage{subcaption}

\title{Gemini Planet Imager Observational Calibrations IX: Least-Squares Inversion Flux Extraction} 

\author{Zachary H. Draper\supit{ab}, Christian Marois\supit{ba}, Schuyler Wolff\supit{cd}, Marshall Perrin\supit{d}, Patrick Ingraham\supit{ef}, Jean-Baptiste Ruffio\supit{gh}, Fredrik T. Rantakyr\"{o}\supit{i}, Markus Hartung\supit{i}, Stephen J. Goodsell\supit{i}, with the GPI team.
\skiplinehalf
\supit{a}University of Victoria, 3800 Finnerty Rd, Victoria, BC, V8P 5C2, Canada; \\
\supit{b}National Research Council of Canada Herzberg, 5071 West Saanich Road, Victoria, BC V8X 4M6, Canada; \\
\supit{c}Physics \& Astronomy Department, Johns Hopkins University, Baltimore MD, 21218, USA;\\
\supit{d}Space Telescope Science Institute, 3700 San Martin Drive, Baltimore MD 21218 USA;\\
\supit{e}Kavli Institute for Particle Astrophysics and Cosmology, Stanford University, Stanford, CA 94305, USA;\\
\supit{f}Department de Physique, Universit\'{e} de Montr{\'e}al, Montr\'eal QC H3C 3J7, Canada;\\
\supit{g}Institute Superieur de l'Aeronautique et de l'Espace, Toulouse, France;\\
\supit{h}SETI Institute, Carl Sagan Center, 189 Bernardo Avenue, Mountain View, CA 94043, USA; \\
\supit{i}Gemini Observatory, Casilla 603, La Serena, Chile; \\
}

\authorinfo{Further author information: (Send correspondence to Z.H.D.)\\Z.H.D.: E-mail: zhd@uvic.ca}

\begin{document} 
  \maketitle 

\begin{abstract}
The Gemini Planet Imager (GPI) is an instrument designed to directly image planets and circumstellar disks from 0.9 to 2.5 microns (the \textit{YJHK} infrared bands) using high contrast adaptive optics with a lenslet-based integral field spectrograph. We develop an extraction algorithm based on a least-squares method to disentangle the spectra and systematic noise contributions simultaneously. We utilize two approaches to adjust for the effect of flexure of the GPI optics which move the position of light incident on the detector. The first method is to iterate the extraction to achieve minimum residual and the second is to cross-correlate the detector image with a model image in iterative extraction steps to determine an offset.  Thus far, this process has made clear qualitative improvements to the cube extraction by reducing the Moir\'{e} pattern. There are also improvements to the automated routines for finding flexure offsets which are reliable to with $\sim0.5$ pixel accuracy compared to pixel accuracy prior.  Further testing and optimization will follow before implementation into the GPI pipeline.
\end{abstract}



\section{INTRODUCTION}
\label{sec:intro}  

The integral field unit on the Gemini Planet Imager (GPI) is a lenslet array based integral field spectrograph (IFS), which disperses light incident on the detector into microspectra over a 2.7$\times$2.7 arcsecond wide field of view (FOV) \cite{JC12,JL14}.  These microspectra are critically sampled on the detector (except in the \textit{Y}-band where they are undersampled) by the lenslet array and offer low resolution spectroscopy across the FOV, creating a datacube of images at varying wavelengths.  The FWHM or line spread function of a single lenslet is $\sim$1.2 pixels \cite{PI14}.  An arc lamp is used to determine the location of these microspectra and establish the wavelength solution.  In quicklook algorithms of the GPI pipeline, rectangular apertures are centered along the spectra and used to determine the flux \cite{MP14,M10}.  Unfortunately, due to non-repeatable flexure in the instrument, the position of the microspectra may be offset from the expected position determined by the wavelength calibration taken at a different orientation during observations. This causes reduced signal-to-noise, inaccurate wavelength calibrations, and contamination of flux into neighboring lenslets.\\

The GPI pipeline currently uses a rectangular aperture method is not an optimal estimator of the flux, because it does not weight the signal in each pixel correctly. The rectangle method also introduces systematic noise effects where a ``checkerboard'' pattern appears within cube slices.  This is attributed to the fact that, due to the regular spacing of the lenslet array, the spectra may fall in either the center or at the edge of a pixel while the extraction is centered only on whole pixels.  This creates an aliasing effect, an alternating pattern of increased and decreased flux between lenslets of the data cube. In addition, bad pixels can lead to pixelization in the data cube at different slices which then require interpolation.\\

\begin{figure}[b]
\begin{center}
\begin{tabular}{c}

\begin{subfigure}[b]{0.6\textwidth}
				\includegraphics[width=\textwidth]{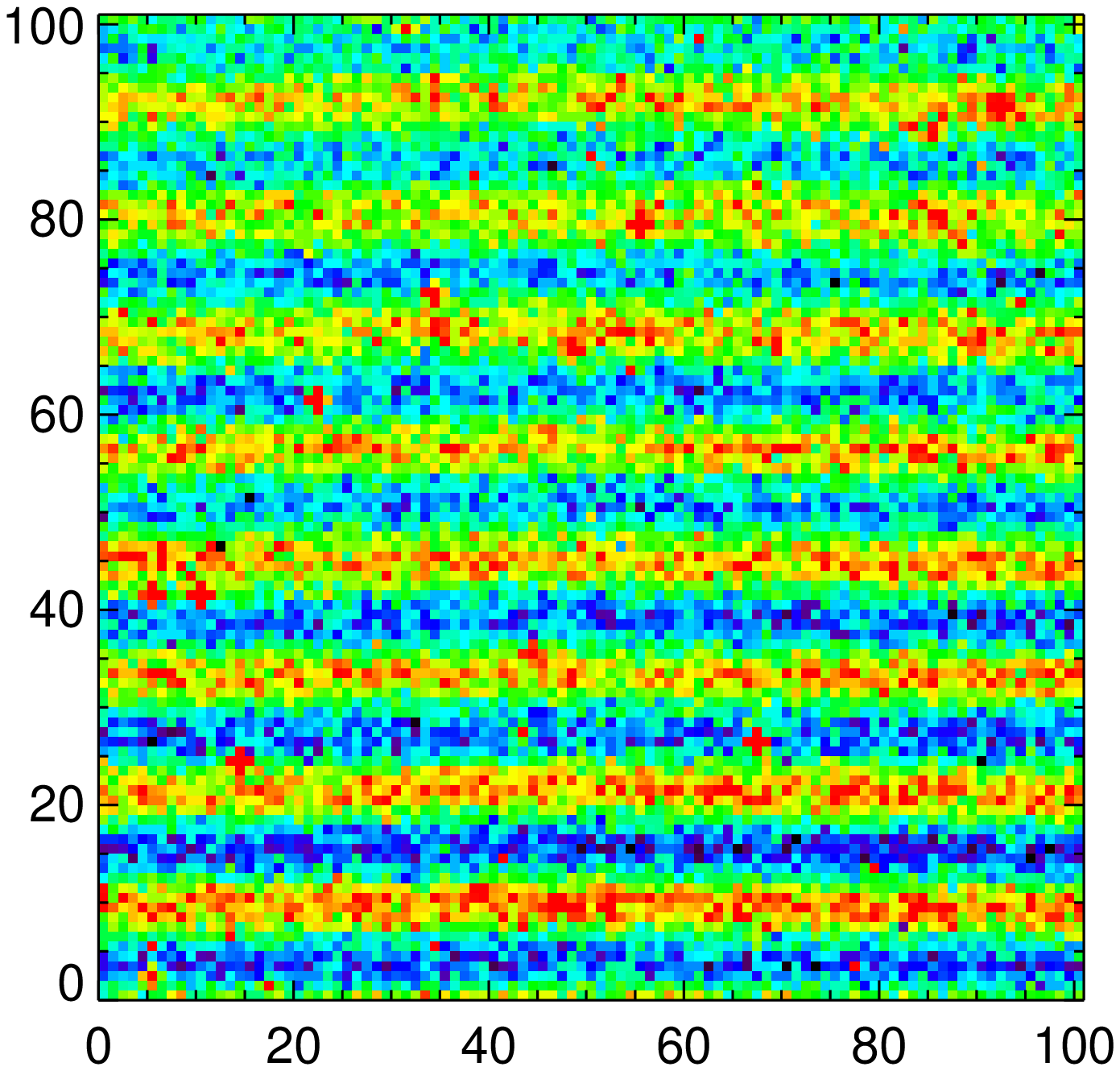}
\end{subfigure}

\begin{subfigure}[t]{0.05\textwidth}
                \includegraphics[trim = 0mm -10mm 0mm 0mm,clip,width=\textwidth]{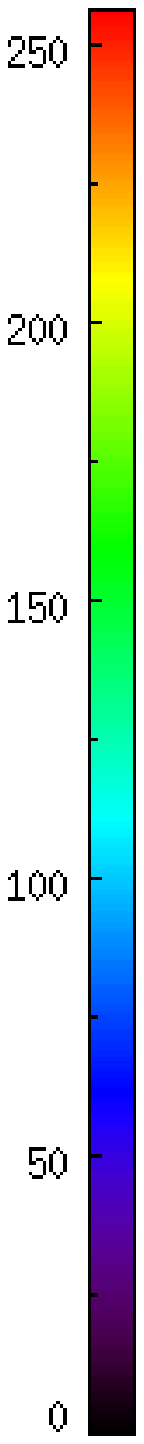}      
\end{subfigure}

\end{tabular}
\end{center}
\caption[] 
{ \label{fig:micphone} 
Sub-section of the detector taken from a dark exposure showing the standing wave pattern induced by the vibrations in GPI.  The x-y axes are in pixels and the image is linearly scaled to minimum and maximum values to enhance the wave pattern.  This constitutes a noise contribution which can be modeled and removed during the extraction process.
}
\end{figure}

An additional noise source is induced by vibrations from the cyrocoolers, which are noticeable in short exposures as a standing wave pattern in the detector (see Figure \ref{fig:micphone}).  This noise contamination on the detector is referred to as microphonic noise (at 60 Hz and harmonics) and can be reasonably modeled and decorrelated with the least squares approach to provide a more accurate datacube extraction.\\

To resolve these issues a more sophisticated approach of PSF extraction is promising to both optimally extract the flux from the detector and adjust for flexure between the wavelength calibration and the science images. High-resolution PSFs are generated using an Anderson and King method which uses under-sampled point sources in combination \cite{AK00,PI14}.  The existing wavelength calibrations are used as a starting point and given an additional global offset to account for the flexure induced since the calibration was made \cite{SW14}.  The flux is then extracted using an inversion method to get flux as a function of wavelength and minimize contamination from neighboring spectra and noise sources (outlined in Section \ref{sec:matrix} and \ref{sec:refimg}).  The flexure offset is found either through an iterative solver or a modeling and cross-correlation routine (outlined in Section \ref{sec:flex}).  \\

\section{REFERENCE IMAGES}
\label{sec:refimg}

In order to run the flux extraction process, we first need to generate reference images for the signal and noise components of the detector images which we can adequately model.  The reference images serve as the model parameters for fitting the detector image. A subsection of the detector image is selected around the microspectra of interest and includes immediate neighbors which may overlap. Correction for the microphonics requires processing the science image of interest, outlined in the following section. The PSF images for signal extraction of the spectra and polarization spots require calibration images which are determined by methods described by other articles in the Gemini Planet Imager Observational Calibrations series. For example see Ingraham, Wolff, these proceedings \cite{PI14,SW14}.

\subsection{MICROPHONICS}
\label{sec:micro}  

In order to isolate microphonic noise, a 64x64 pixel subsection of the CCD detector is selected from within a single amplifier band.  The section is median-filtered to remove the signal contribution to the image.  A 2D Fourier transform is used to select the frequencies of the microphonics pattern closest to the frequency of vibration.  Since the microphonics are predominately in the vertical direction, only vertical frequency components are used.  This is confirmed by the 2D Fourier power spectrum.  Then several images are made by generating sine and cosine images with those frequencies which have the highest power from the Fourier transform. This approach is an alternative to the destriping algorithm (See Paper II) in the GPI pipeline because the least square extraction algorithm allows for coherent noise contributions to be modeled and decorrelated from the signal simultaneously \cite{PI14B}.

\begin{figure}[b]
\begin{center}
\begin{tabular}{c}

\begin{subfigure}[b]{0.33\textwidth}
                \includegraphics[trim = 40mm 0mm 40mm 0mm,clip,width=\textwidth]{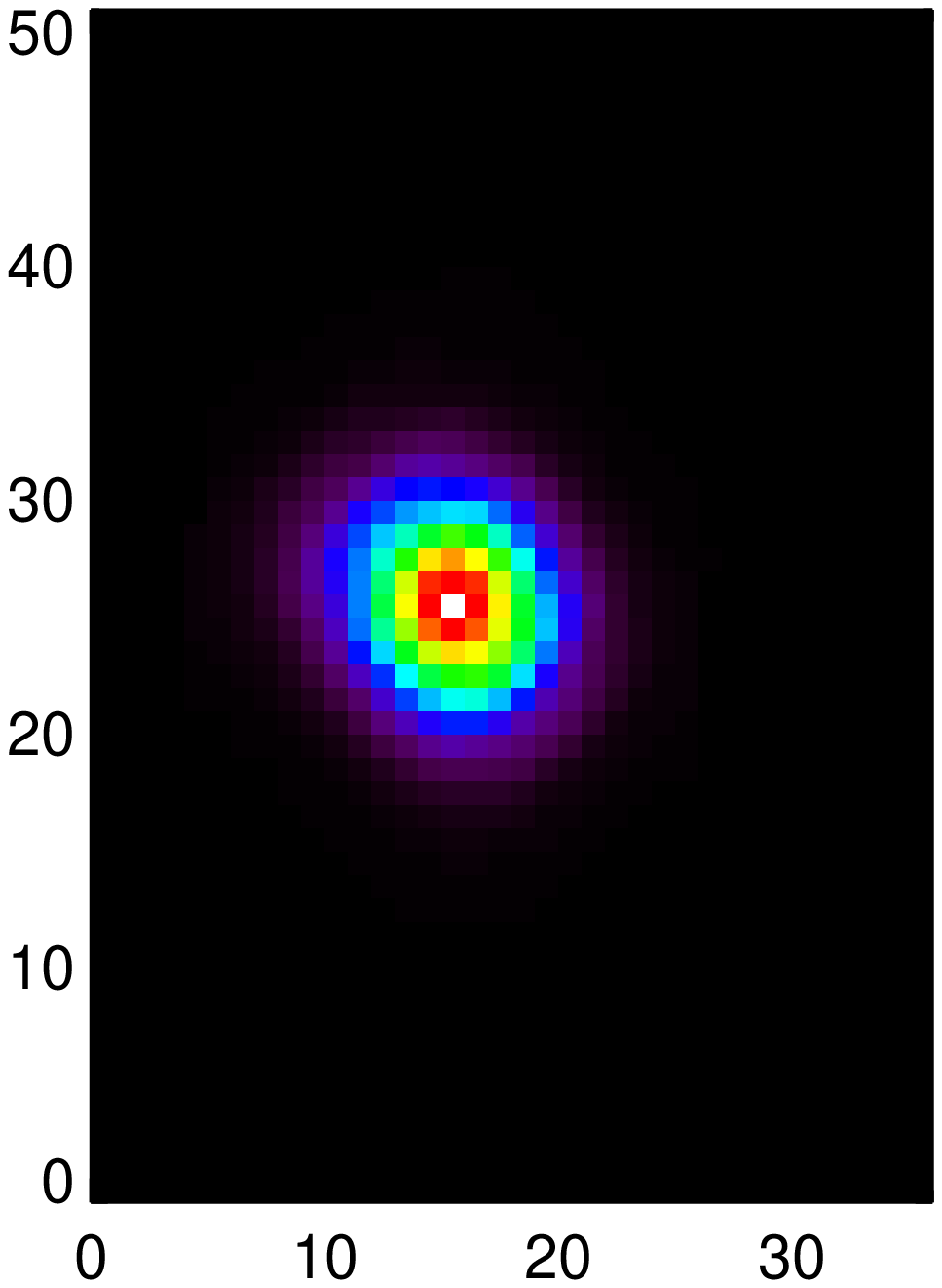}
                \caption{High Resolution Empirical PSF}
                \label{fig:high_res}
\end{subfigure}

\begin{subfigure}[b]{0.33\textwidth}
                \includegraphics[trim = 40mm 0mm 40mm 0mm,clip,width=\textwidth]{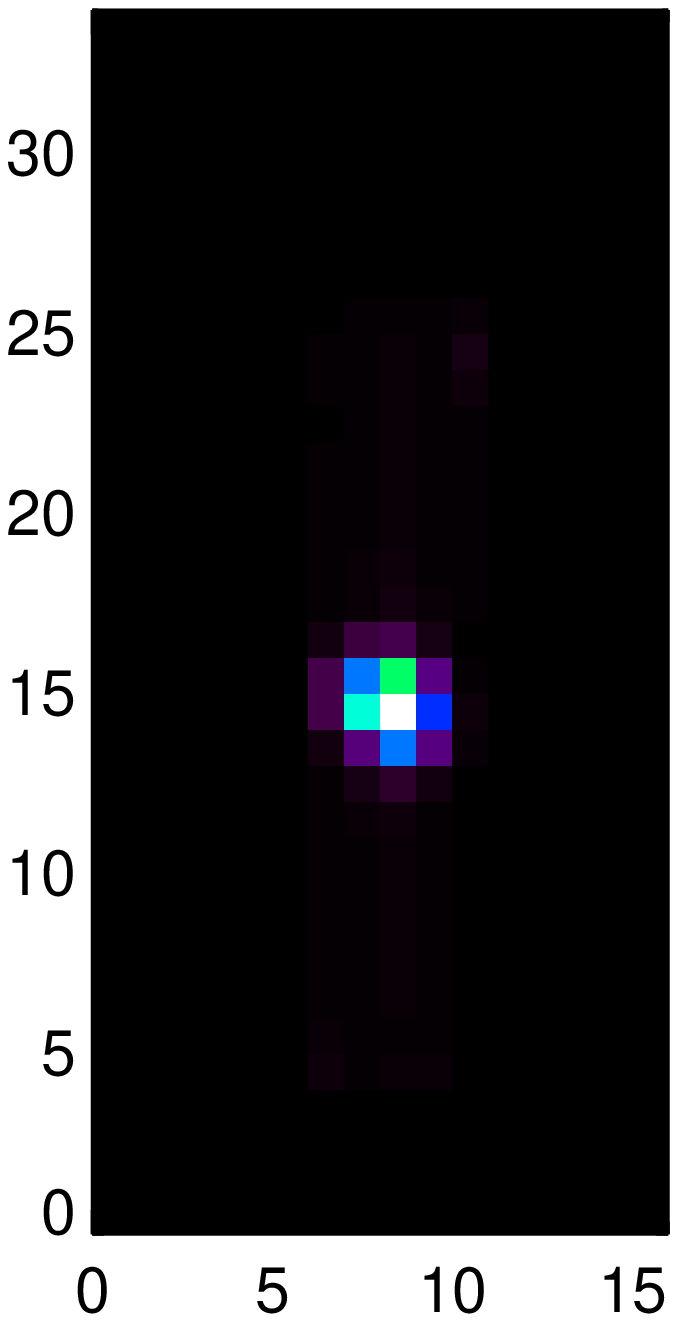}
                \caption{Detector Resolution PSF}
                \label{fig:d_res}
\end{subfigure}

\end{tabular}
\end{center}
\caption{\textbf{(a)} High resolution PSF generated using the method of Paper III with emission line spectra \cite{PI14}. \textbf{(b)} PSF binned to detector resolution and positioned on a microspectra  within a subset image using a GPI wavelength calibration. This is a reference image for a single wavelength on the central microspectra}
\end{figure}

\subsection{HIGH-RESOLUTION PSF}
\label{sec:highres} 

A high-resolution model for the microlenslet PSF is derived using a method developed originally for HST WFPC2 by Anderson and King \cite{AK00,PI14}.  The empirical model of the PSF is generated using arc lamp spectra from GPI at various sub-pixel positions (see Figure \ref{fig:high_res}).  For more details on how this algorithm is implemented for GPI see Ingraham et al., these proceedings \cite{PI14}.  For each sub-image, a high-resolution PSF is selected and then interpolated with a bilinear interpolation to a subpixel position and grid appropriate to the subset image (see Figure \ref{fig:d_res}). \\  

The subpixel position at which to place the reference PSF image within the spectra is determined from the wavelength calibration determined by emission lamp spectra (see Wolff at al., these proceedings) \cite{SW14}.  The PSFs are separated by the resolution limit of $\sim$2 pixels to preserve the stability of the matrix inversion, so that no two reference images are extremely correlated which would cause oscillating positive and negative solutions.  PSFs are placed on the spectra of interest centered in the subset images as well as the neighboring spectra.  This is done to remove contamination from neighboring lenslets within the image.  For example, near the edges of the detector in all infrared bands, spectra are tilted at increasingly large angles due to the properties of the refractive optics within the IFS and start to blend with neighboring spectra.  In the K-bands (GPI has two filters to cover the K-band referred to as K1 \& K2), a section down the middle of the detector has spectra touching end to end from each lenslet.  Extracting these spectra simultaneously allows spectra to be decorrelated to produce a cleaner data cube than the rectangular aperture algorithm (see Figures \ref{fig:datacubes1} and \ref{fig:datacubes2}). \\

In polarization mode, light from each lenslet is split into two orthogonal polarization states  via a Wollaston prism \cite{M10}.  The spots are sufficiently separated and uncorrelated with other spots. This means they only require a single PSF for flux extraction.  High resolution versions of the polarization mode spots are made separately using unpolarized flat field images with the same algorithm used for spectral mode.  Due to chromatic aberrations, polarization mode lenslet PSFs are sufficiently different from spectral mode PSFs to warrant the use of different PSFs between the two modes.  \\

\begin{figure}[h]
\begin{center}
\begin{tabular}{c}

\begin{subfigure}[b]{0.23\textwidth}
                \includegraphics[trim = 59mm 0mm 59mm 0mm,clip,width=\textwidth]{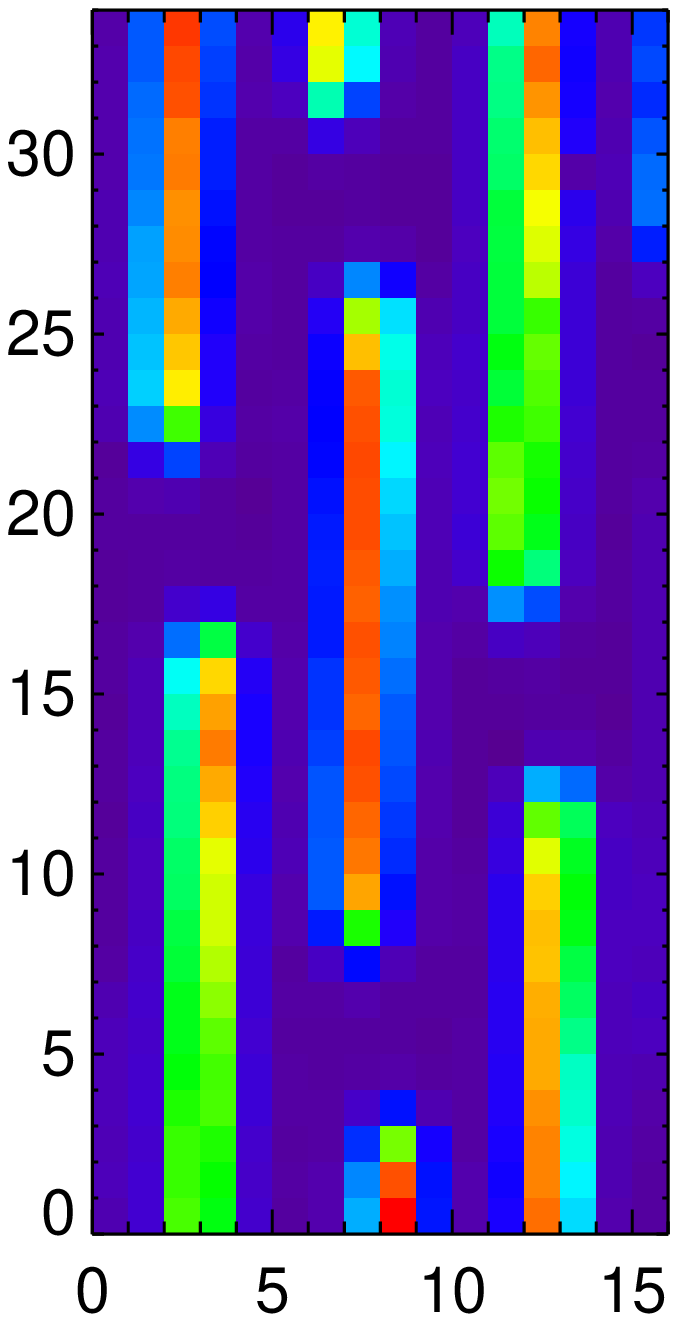}
                \caption{Data}

\end{subfigure}

\begin{subfigure}[b]{0.23\textwidth}
                \includegraphics[trim = 59mm 0mm 59mm 0mm,clip,width=\textwidth]{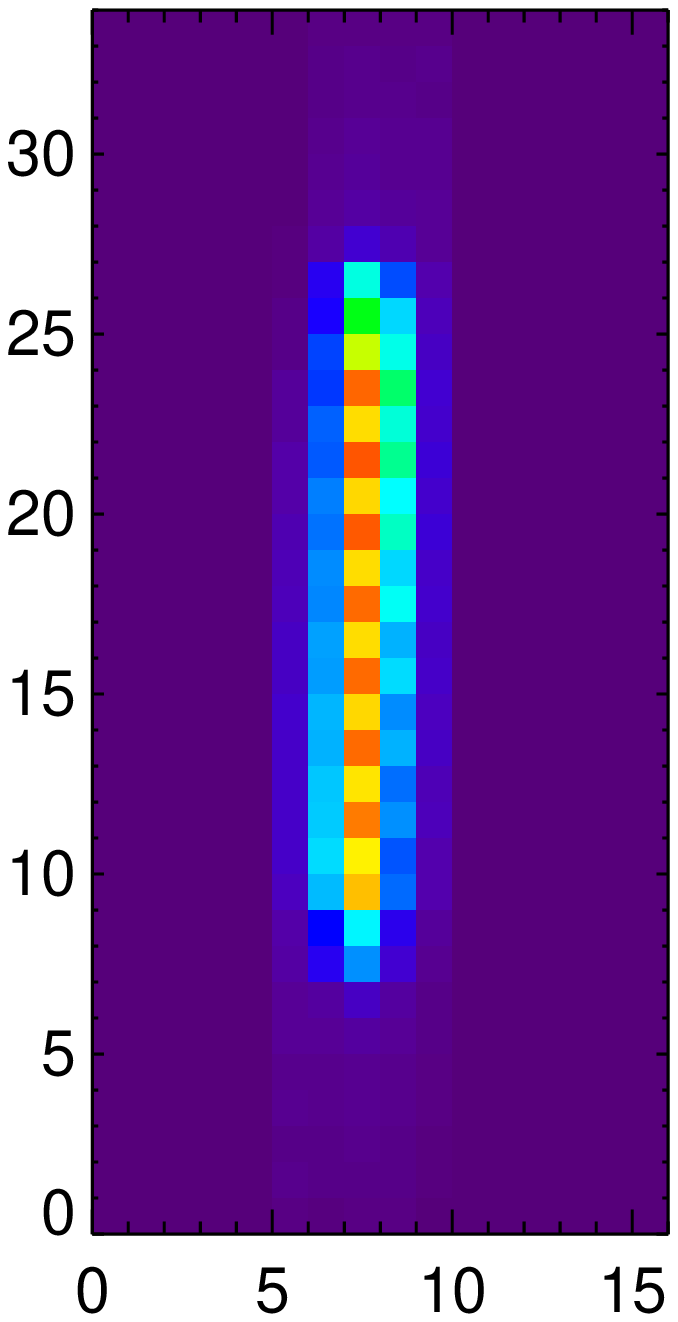}
                \caption{Extracted Flux}

\end{subfigure}

\begin{subfigure}[b]{0.23\textwidth}
                \includegraphics[trim = 59mm 0mm 59mm 0mm,clip,width=\textwidth]{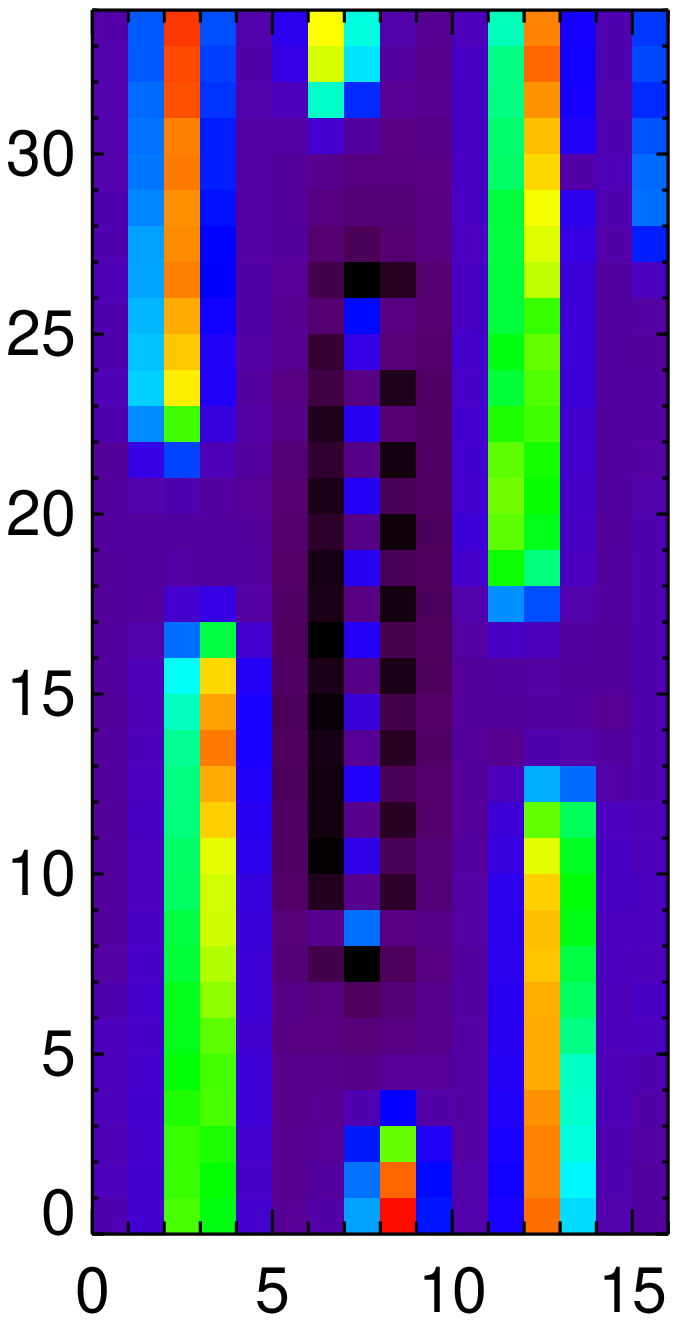}
                \caption{Extracted Residual}

\end{subfigure}

\begin{subfigure}[b]{0.23\textwidth}
                \includegraphics[trim = 59mm 0mm 59mm 0mm,clip,width=\textwidth]{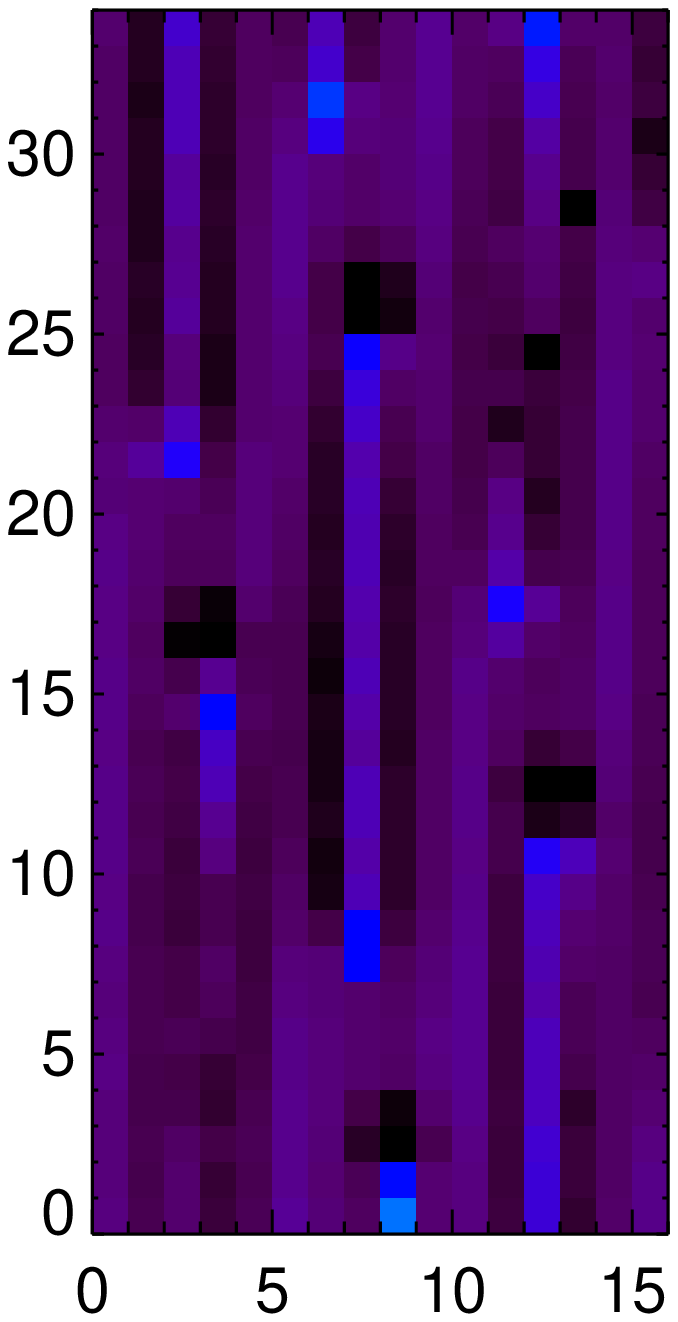}
                \caption{Modeled Residual}
\end{subfigure}

\begin{subfigure}[t]{0.05\textwidth}
                \includegraphics[trim = 0mm -20mm 0mm 0mm,clip,width=\textwidth]{tst.eps}
\end{subfigure}

\end{tabular}
\end{center}
\caption[example] 
{ \label{fig:example} 
Spectral extraction using least squares PSF method. The x an y axes are in pixels and share a linear color scaled derived from the minimum and maximum counts in the original detector image.  \textbf{(a)}: Subsection of the raw detector image centered on a microspectra.  \textbf{(b)}: Reconstructed spectra using the PSFs extracted. \textbf{(c)}: Residual for the detector image minus the extracted spectrum.  Note that the oscillating pattern in the y-direction is due to the resolution limit, which prevents continuous spacing between PSFs for the inversion method.  This necessitates subpixel dithering to complete the interpolation of flux between each PSF in the final GPI datacube. \textbf{(d)}: A full residual using the spectrum gained from dithering the extraction, modeling the detector at 0.1 pixel PSF separation, and subtracting from the data.}
\end{figure} 

\section{INVERSION ALGORITHM}
\label{sec:matrix} 

The algorithm for flux extraction follows a linear algebra approach. The least-squares solution is found using a basis set formed from a system of known reference images. The basis set that can then be used to model the data image being fit.  Similar applications were used in other astronomical image processing pipelines for PSF subtraction \cite{DL07,CM10}. This approach involves the inversion of a correlation matrix of the reference images in order to determine each individual reference image's contribution to the data image.  

Using the definition of a least square we define D to be the data image and M to be the model image which is fit to the data.

\begin{equation}
\label{eq:ls1}
\chi^{2} = (D - M)^{2}
\end{equation}

The model image is the product of a set reference images (outlined in the previous section) and a coefficient vector equivalent to the flux within a given PSF or power of a microphonics image. We define the set of reference images as $A_{k}=\{R_{0},...,R_{k}\}$ and the coefficient vector as $\vec{f}_{k}$.

\begin{equation}
\label{eq:ls2}
\chi^{2} = (D - \sum_{k} \vec{f}_{k} A_{k})^{2}
\end{equation}

Taking the derivative and setting it to zero we find the least square estimator of the vector $\vec{f}_{k}$, where $A_{j}$ is an identical set to $A_{k}$.

\begin{equation}
\label{eq:ls3}
\frac{\partial \chi^{2}}{\partial \vec{f}_{k}} = 2 \sum_{j} A_{j} (D - \sum_{k} \vec{f}_{k} A_{k}) = 0
\end{equation}

\begin{equation}
\label{eq:ls4}
\sum_{j} A_{j} D = \sum_{j} A_{j} \sum_{k} \vec{f}_{k} A_{k}
\end{equation}

Rearranging and simplifying terms we get Equation \ref{eq:vector_i}, where \textbf{C} is a correlation matrix of the reference images with themselves and $\vec{v}$ is a vector of the flux from each reference multiplied by the data image.

\begin{equation}
\label{eq:ls4}
\left( \sum_{j} A_{j} D \right) = \left( \sum_{j} A_{j} \sum_{k} A_{k} \right) \times \vec{f}_{k}
\end{equation}

\begin{equation}
\label{eq:vector_i}
\vec{v} = \textbf{C} \times \vec{f}_{k}
\end{equation}

First, the correlation matrix (\textbf{C}) is generated by taking the set of reference images, $R_{0}$ through $R_{k}$, and element-wise multiplying with an identical set of reference images, $R_{0}$ through $R_{j}$, where \textit{j} and \textit{k} are the reference image numbers. The product of the two images is element-wise summed to get the relative degree of correlation between the two reference images (See Equation \ref{eq:corr_mat}).  The result is a square matrix where each element of the correlation matrix ($c_{jk}$) is the sum of the product of reference image number \textit{j} with image number \textit{k}.  Each reference PSF image is normalized to a constant before multiplication to give equal weight between PSFs.  This results in a square matrix whose size depends on the number of reference images required for the subset data image.

\begin{equation}
\label{eq:corr_mat}
\textbf{C} = \left[ 
	\begin{array}
		{ccc} c_{00} & \cdots & c_{j0} \\ \vdots & \ddots & \vdots \\ c_{0k} & \cdots & c_{jk}
	\end{array}
\right]
; \rm{where} \; \textit{c}_{jk} = \sum ( R_{j} \circ R_{k} )
\end{equation}

Secondly, each reference image is then multiplied with the subset data image (D) and summed to yield a vector of the detector counts within each reference image.

\begin{equation}
\label{eq:vector}
\vec{v} = \left[ 
	\begin{array}
		{ccc} v_{0} \\
				\vdots \\
					 v_{k} \\
	\end{array}
\right]
; \rm{where} \; \vec{\textit{v}} = \sum (D \circ R_{k})
\end{equation}

The third step is to invert the correlation matrix through standard Gaussian elimination with IDL INVERT and solve for $\vec{f}_{k}$.  Non-negative matrix inversions were also tested. They allowed the PSFs to be more closely spaced but increased the computation time and still required a dithering approach to provide a more complete interpolation of the underlying spectrum.  PSFs which are sufficiently spaced do not tend to reach negative values, which are considered unreal systematic noise as the detector should only have positive flux contributions.  Singular-value decomposition matrix inversion was also tested but again provided no substantial improvement over the simpler method. The inverse correlation matrix is multiplied by the flux vector to get the reference image coefficients within a least squares minimum residual (see Figure \ref{fig:example}). For reference images of a PSF the coefficient is equivalent to the monochromatic flux at the wavelength determined by the wavelength calibration.

\begin{equation}
\label{eq:vector}
\vec{f}_{k} = \textbf{C}^{-1} \times \vec{v}
\end{equation}

This process is repeated after shifting the PSF locations by a two thirds of resolution limit to either side of the accepted wavelength calibration along the dispersion axis to build up flux contributions at other wavelength values.  A GPI cube is then interpolated at a uniform wavelength separation at each lenslet of the array.

\begin{figure}[h]
\begin{center}
\begin{tabular}{c}

\begin{subfigure}[b]{0.23\textwidth}
                \includegraphics[trim = 59mm 0mm 59mm 0mm,clip,width=\textwidth]{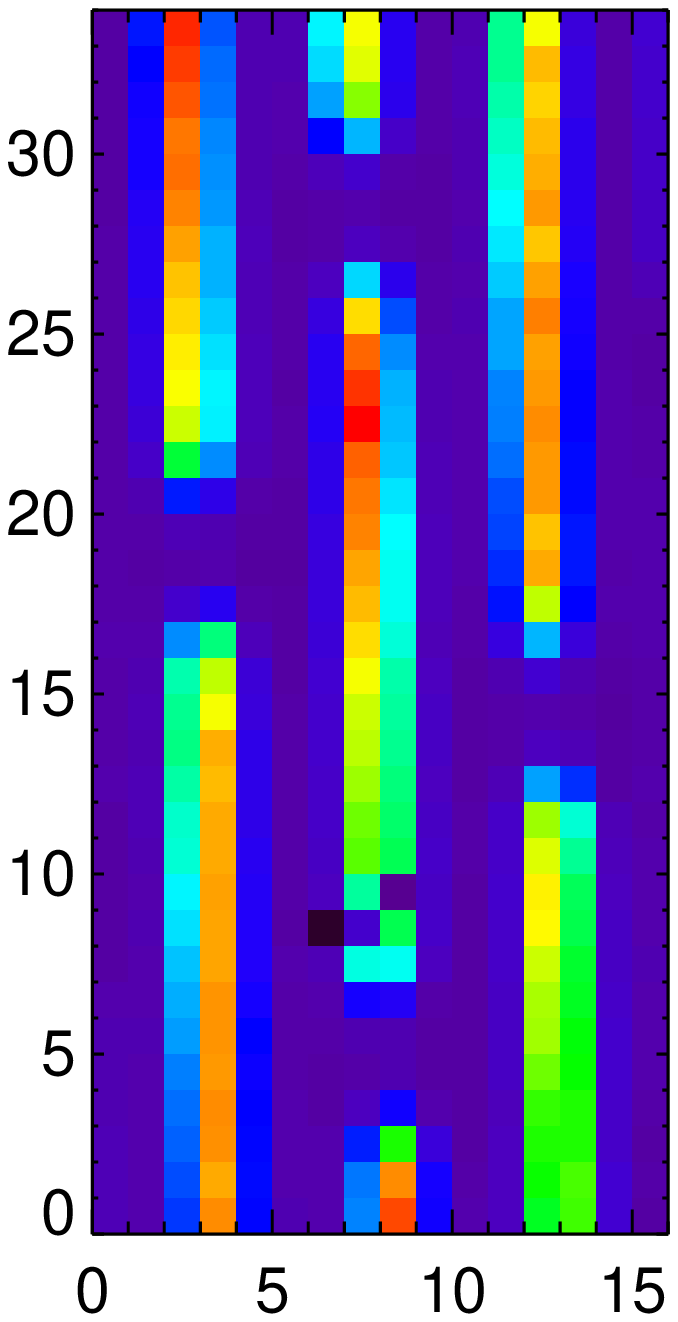}
                \caption{Data}

\end{subfigure}

\begin{subfigure}[b]{0.23\textwidth}
                \includegraphics[trim = 59mm 0mm 59mm 0mm,clip,width=\textwidth]{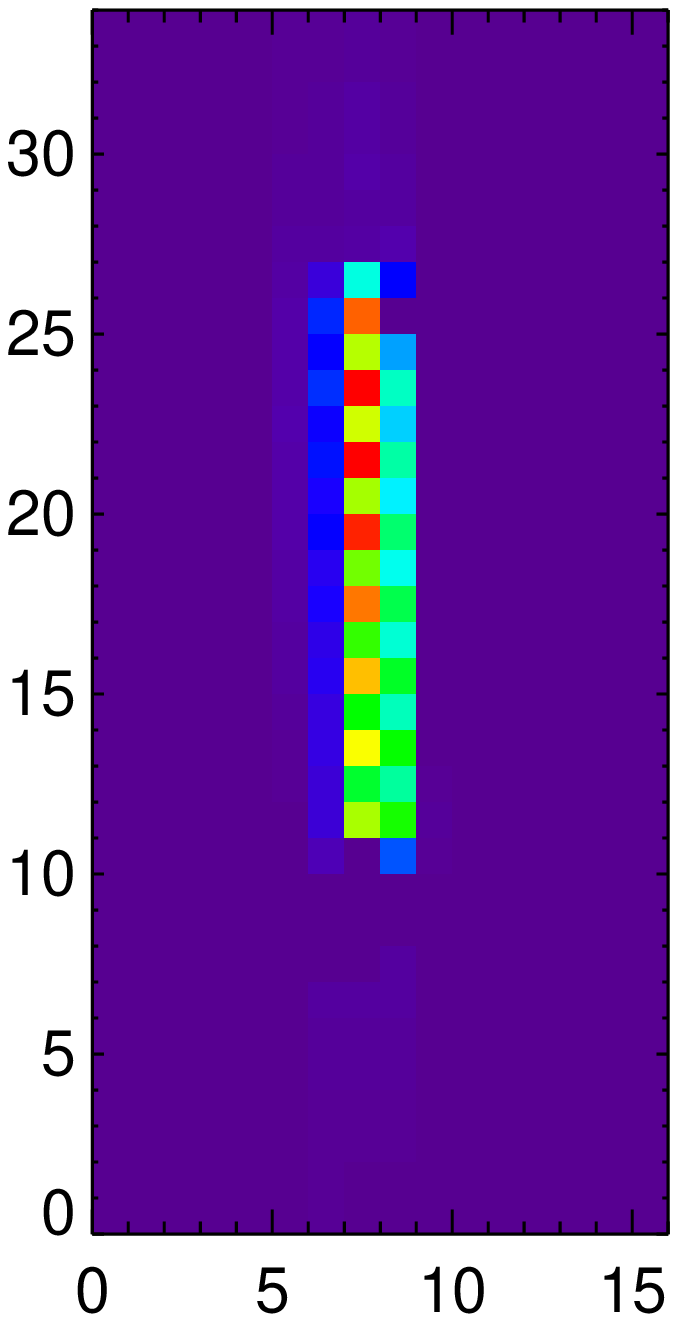}
                \caption{Extracted Flux}

\end{subfigure}

\begin{subfigure}[b]{0.23\textwidth}
                \includegraphics[trim = 59mm 0mm 59mm 0mm,clip,width=\textwidth]{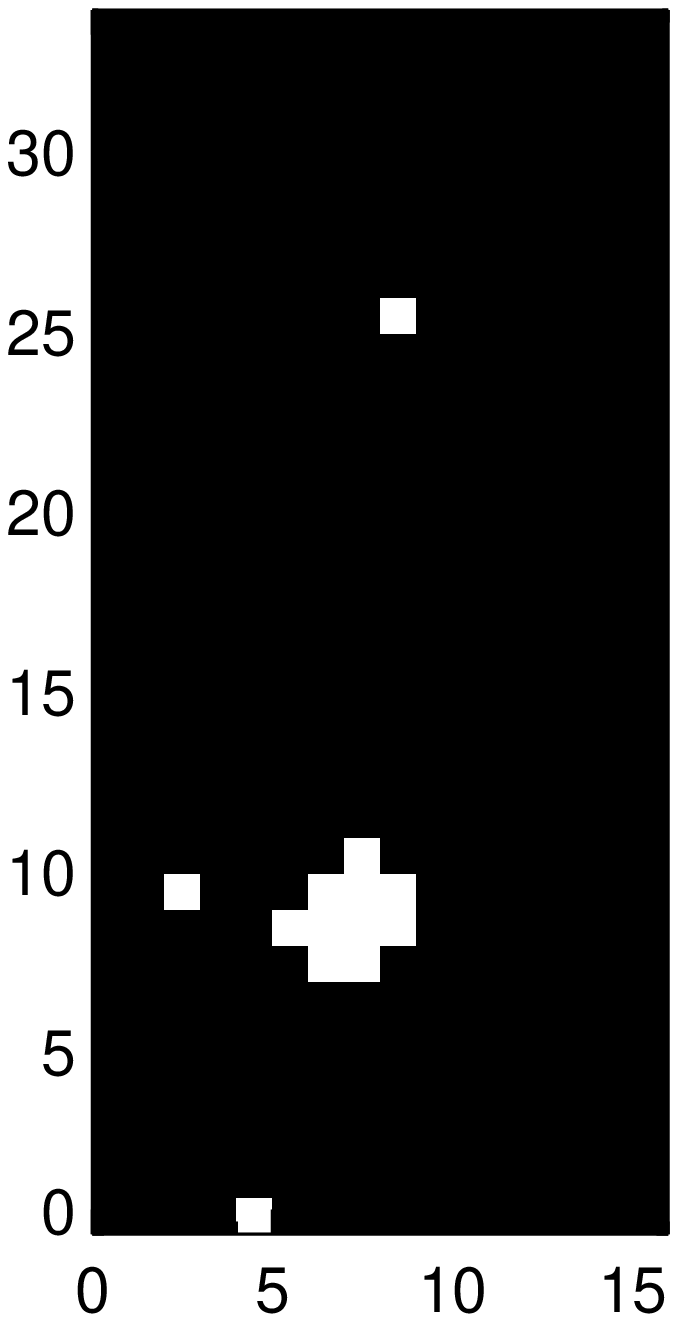}
                \caption{Bad Pixel Mask}

\end{subfigure}

\begin{subfigure}[b]{0.23\textwidth}
                \includegraphics[trim = 59mm 0mm 59mm 0mm,clip,width=\textwidth]{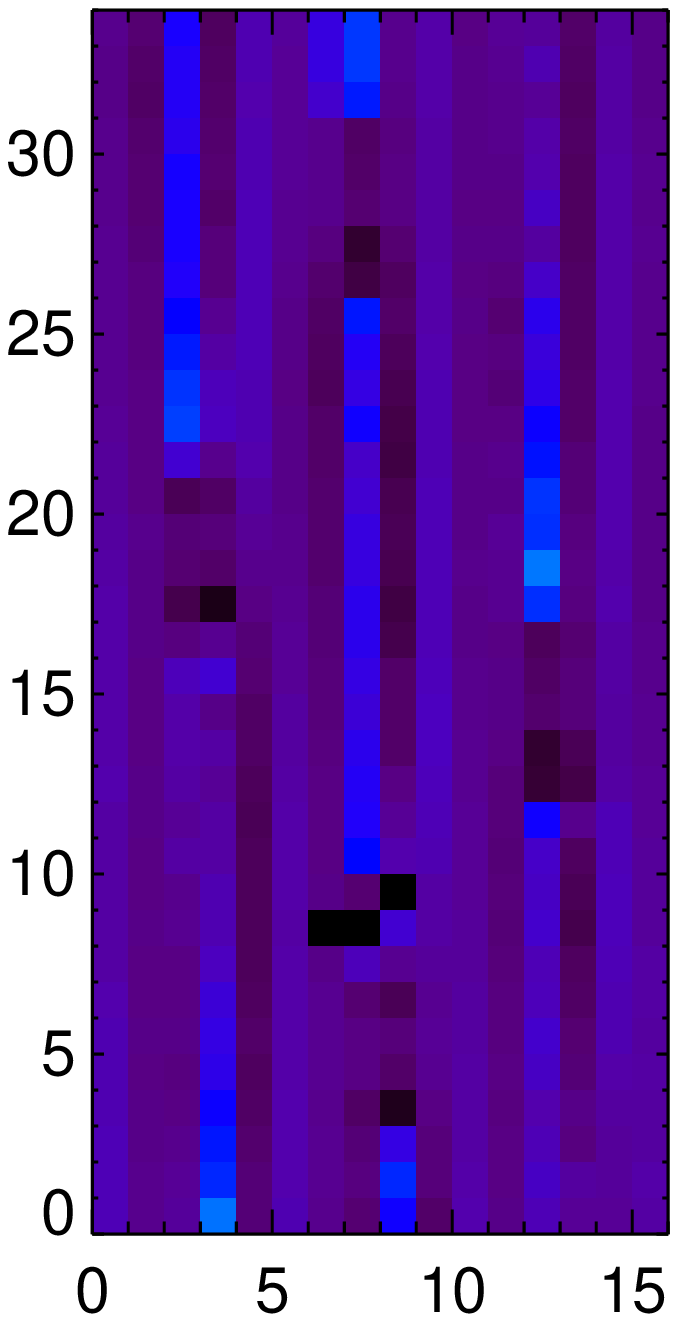}
                \caption{Modeled Residual}

\end{subfigure}

\begin{subfigure}[t]{0.05\textwidth}
                \includegraphics[trim = 0mm -20mm 0mm 0mm,clip,width=\textwidth]{tst.eps}      
\end{subfigure}

\end{tabular}
\end{center}
\caption[] 
{\label{fig:bpx} 
Another spectral extraction showing bad pixel masking. XY axis are in pixels and linearly scaled to the same maximum and minimum values. \textbf{(a)}: Raw Detector, \textbf{(b)}: Extracted spectra, and \textbf{(c)}: Bad pixel mask applied to weight extraction.  Note the gap in the extracted spectrum image requires interpolating across the gap in the spectrum as a function of wavelength. \textbf{(d)}: Shows the full residual after spectral interpolation.}
\end{figure} 

\subsection{BAD PIXEL MASKING}
\label{sec:badpixel} 

It is also an option to isolate bad pixels during the flux extraction process.  Bad pixels are identified through an accumulation of dark images.  For more information see Ingraham et al., these proceedings \cite{PI14B}. Once the reference images are made, the bad pixels within an extracted subset image are set to zero through element wise multiplication of the bad pixel mask (B). This is done before the correlation matrix (\textbf{C}) and flux vector ($\vec{v}$) are calculated so that they do not contribute to the extracted flux value during the least squares algorithm (see Figure \ref{fig:bpx}, e.g.).

\begin{equation}
\label{eq:bpmask}
R_{masked} = R \circ B
\end{equation}

\section{FLEXURE OFFSETS}
\label{sec:flex} 

GPI is mounted at the Cassegrain focus at Gemini South and therefore experiences a varying gravity vector during an observing sequence.  Flexure between the lenslet array and the detector causes the light to move upwards of 2 pixels in any given direction on the hybrid CMOS detector \cite{MP14}.  The result is that an earlier wavelength solution taken with the telescope at a different elevation wont be directly valid.  In order to adapt to flexure during observations we employ two methods to find the optimal signal extraction, iterative wavelength calibration offsets and 2D cross-correlation.  Primarily the concern is to find a global offset, however there are FOV effects which require localized offsets for each lenslet \cite{SW14}. The iterative solver can find offsets for each lenslet or take a global average while the cross correlation method finds only the global offset.  Both of these approaches work in two different coordinate regimes, detector xy-shifts and dispersion coordinates of angle, perpendicular, and parallel shifts.  This is due to degeneracies discussed below.  

\subsection{ITERATIVE SOLVER}
\label{sec:iter} 

Using the IDL AMOEBA downhill simplex method, the wavelength calibration is given three offset parameters to offset in order to find the minimum residual.  The perpendicular shift is an offset to sidestep the spectra into place. The angle parameter adjusts the angle of the dispersion axis.  Finally, the parallel offset adjusts the position along the spectra.  The biggest problem with this method is that the offset along the dispersion axis is not well constrained without spectral information because the extraction algorithm can find a minimum residual with some PSFs falling outside the spectrum.  While shifting the extraction, PSFs are free to "slide" off the end of the band such that one PSF at the end of the wavelength range is extracting detector noise, especially in low SNR regime.  Effectively the minimum residual does not represent the wavelength solution of the light accurately.  Another option which is implemented is to do a two parameter offset in xy-detector coordinates, but because the y-axis is predominately the dispersion axis, there is a similar problem.  This is partially mitigated by using spectra on the edge of the detector which have spectra at different dispersion angles, such that they are not completely vertical. \\

	Another problem arising with this method is that it requires solving for the reference images and inversion algorithm $\sim$10-20 times for a 2 pixel range of offsets at a 0.1 pixel convergence. This drastically increases the computation time over using a single extraction with a global offset. Furthermore, if the minimum residual is not constrained to common value, the wavelength solution will vary across the FOV such that flux at one end of the image is from a different wavelength of light then the other side.

\subsection{2D CROSS-CORRELATION}
\label{sec:xcorr} 

\begin{figure}
\begin{center}
\begin{tabular}{c}

\begin{subfigure}[b]{0.25\textwidth}
                \includegraphics[trim = 70mm 0mm 70mm 0mm,clip,width=\textwidth]{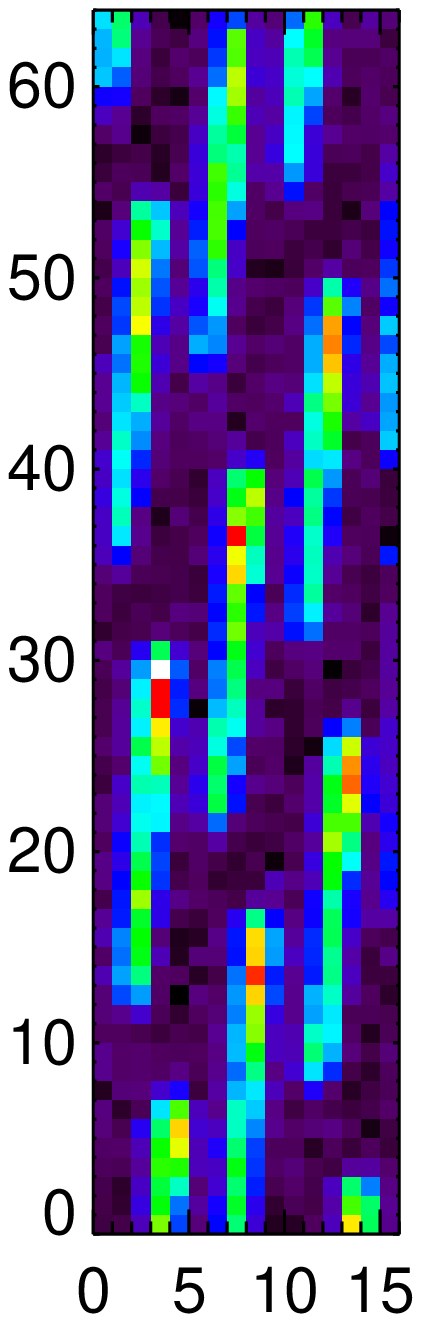}
                \caption{Data}
                \label{fig:data}
\end{subfigure}

\begin{subfigure}[b]{0.25\textwidth}
                \includegraphics[trim = 70mm 0mm 70mm 0mm,clip,width=\textwidth]{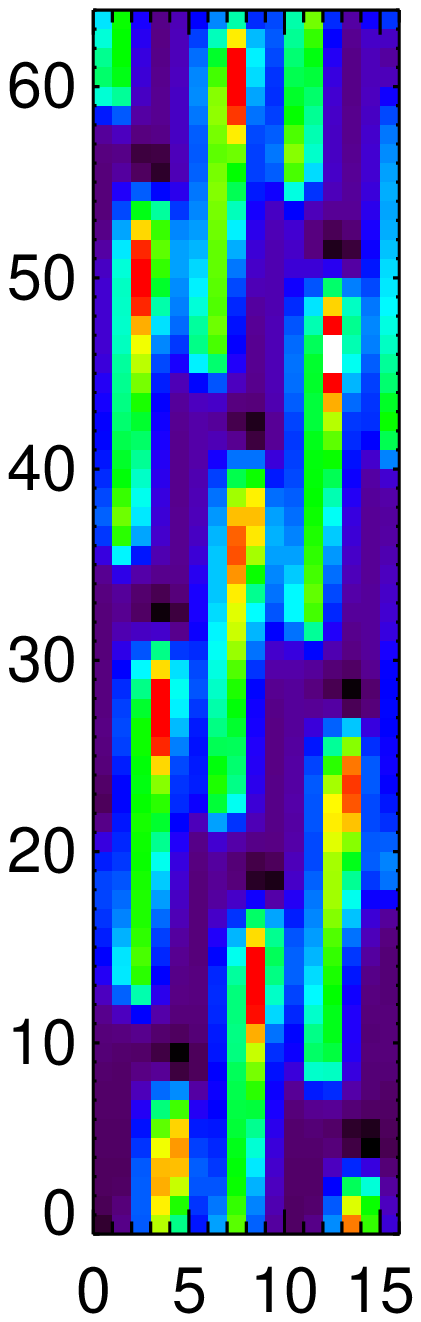}
                \caption{Model}
                \label{fig:model}
\end{subfigure}

\begin{subfigure}[b]{0.25\textwidth}
                \includegraphics[trim = 70mm 0mm 70mm 0mm,clip,width=\textwidth]{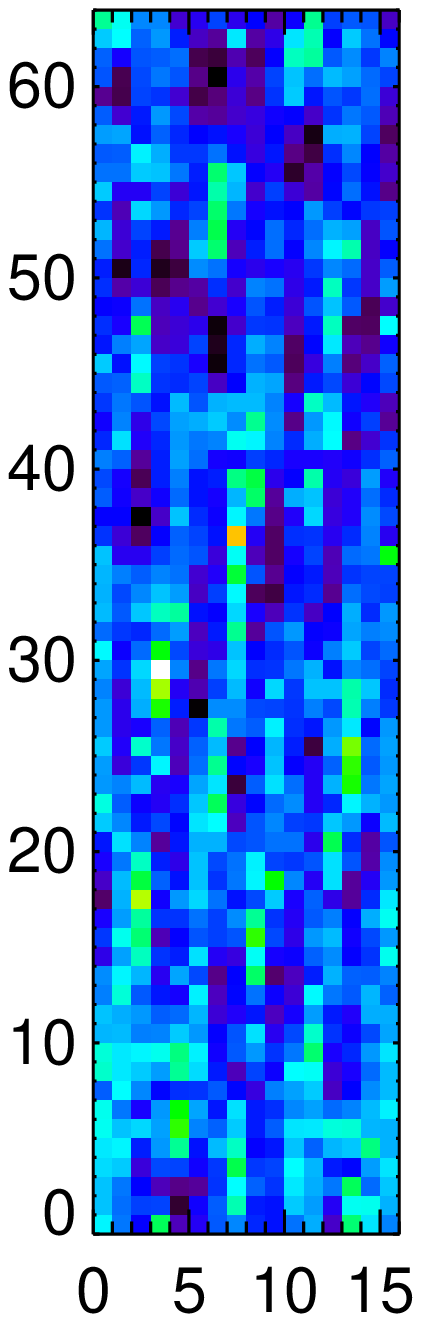}
                \caption{Residual}
                \label{fig:residual}
\end{subfigure}

\begin{subfigure}[t]{0.07\textwidth}
                \includegraphics[trim = 0mm -27mm 0mm 0mm,scale=2,clip,width=\textwidth]{tst.eps}
\end{subfigure}

\end{tabular}
\end{center}
\caption{\textbf{a)} Detector subsection with stellar spectra from standard star. \textbf{b)} Model image of the detector given a dithered spectral extraction from the data image, the microlens PSF seperated at 0.01 pixels, and wavelength calibration adjusted for flexure offset. \textbf{c)} Residual frame of the model after subtraction.}
\end{figure}

Another method which is used to determine the flexure is to cross-correlate a modeled spectra with the detector image (see Figure \ref{fig:data}).  First the extraction algorithm is run to compute a rough spectra of a large subset image, which is either the observing target's stellar flux or sky background.  The extracted spectrum is used to forward model the data (see Figure \ref{fig:model}) to provide an image for cross-correlation.  The peak correlation is found using the maximum total value for the data and the model image given a grid of subpixel offsets through Fourier transforms.  The offsets produced in subpixel detector xy-coordinates can then be used to re-extract the spectrum and iterate the forward modeling and cross-correlation until a convergence in shifts results in a minimal residual (see Figure \ref{fig:residual}).  The entire routine takes less then 5 calls with a 2 pixel range search, which narrows with each iteration, and at $\sim0.05$ pixel convergence.\\

The benefit of this method is that fewer calls to the extraction algorithm are needed, but more computation is spent on generating the model image for the 2D cross-correlation.  In practice, this method takes much less time then the iterative solver: on the order of 3 minutes on a single processor rather than 40 minutes on 15 processors. The dispersion offset (predominately detector y-axis offsets with angle $\approx 0^{\circ}$) is better constrained in this method because the shape of the spectra play a role in the derived offsets.  This is in part due to modeling the edge of the detector where the spectra have been distorted to have a non-zero dispersion angle.  Having spectra at different angles results in a more constrained cross-correlation when used in combination, rather than having a completely repeated, uniform shape.  The cross-correlation can also be run in 1D as a perpendicular shift by combining detector xy-offsets at a fixed angle and converges reliably for large offsets. This does not however represent the full flexure offset as flexure will also cause a shift along the dispersion axis.\\

In the case of the polarization mode, the spots are less complex to match given they are single PSF sources. Typically, one to two iterations of modeling and cross correlating are sufficient to find the offsets to subpixel accuracy.  Also, they do not suffer from the y-offsets being anymore indeterminate then the x-offset due to their point like nature.\\

\begin{table}[h]
\caption{The global average XY-offset due to flexure are found for the following observations. The wavelength calibration is given artificial flexure offsets (XY-initial) to test for convergence with different starting parameters. Offsets are in pixels, while telescope orientation is in degrees.  The telescope elevation is shown to be very close, such that they should have the same flexure offset.  The arc lamp is found to have the true offset by matching the emission line spectra.  The Y-offset does not converge to a single value as well as the X-offset, but taken as an average, are quite close to the accepted values from the Ar Arc lamp.} 
\label{tab:flex}
\begin{center}
\begin{tabular}{|c|cc|cc|c|c|}
\hline
\rule[-1ex]{0pt}{3.5ex} Target (Band) & X-Offset (px) & Y-Offset (px) & X-Initial (px) & Y-Initial (px) & Elevation $(^{\circ})$ \\
\hline
\rule[-1ex]{0pt}{3.5ex} Ar Arc (H)	& \textbf{-0.46} & \textbf{-0.07} & 0 & 0 & 66.1 \\
\hline
\rule[-1ex]{0pt}{3.5ex} PZ Tel (H) & -0.42 & -0.2 & 0 & 0 & 65.4 \\

\rule[-1ex]{0pt}{3.5ex} \vdots & -0.41 & -0.59 & -1 & -0.5 & \vdots \\

\rule[-1ex]{0pt}{3.5ex} \vdots & -0.42 & 0.18 & -1 & 0.5 & \vdots \\

\rule[-1ex]{0pt}{3.5ex} \vdots & -0.42 & 0.12 & 0.5 & 0.5 & \vdots \\	

\rule[-1ex]{0pt}{3.5ex} \vdots & -0.40 & -0.88 & 0.5 & -1 & \vdots \\	
\hline
\rule[-1ex]{0pt}{3.5ex} Mean PZ Tel & \textbf{-0.41} & \textbf{-0.27} & All & All & 65.4  \\	
\hline
							

\end{tabular}
\end{center}
\end{table}

In order to determine how well the routine converges to a true flexure offset, science data is used in conjunction with an arc lamp observation at the same telescope elevation taken just prior to the science target. The modeling and cross correlation is run on the PZ Tel data taken on May 11th with varying initial offsets from the wavelength calibration to test for convergence.  Table \ref{tab:flex} shows the offsets determined by the modeling and cross correlation routine for an Ar Arc lamp taken just prior to an observation of PZ Tel in the H-band.  The XY-offset is found for the arc lamp more precisely by matching the extracted emission lines with their rest wavelength. These offsets are assumed to be the true flexure offset from the wavelength calibration because we know the spectrum precisely. For the science image of PZ Tel, the X-offset converges well (-0.46 vs -0.41 pixels), but the Y-offset still has some variation due to the changing spectral shape (-0.07 vs -0.27 pixels). Using stricter convergence criteria tends to make the solution divergent and may be a limitation on how well the spectra can be extracted and subsequently modeled.\\

\section{Conclusion}

\begin{figure}
\begin{center}
\begin{tabular}{c}

\begin{subfigure}[b]{0.48\textwidth}
                \includegraphics[trim = 15mm 8mm 17mm 5mm,clip,width=\textwidth]{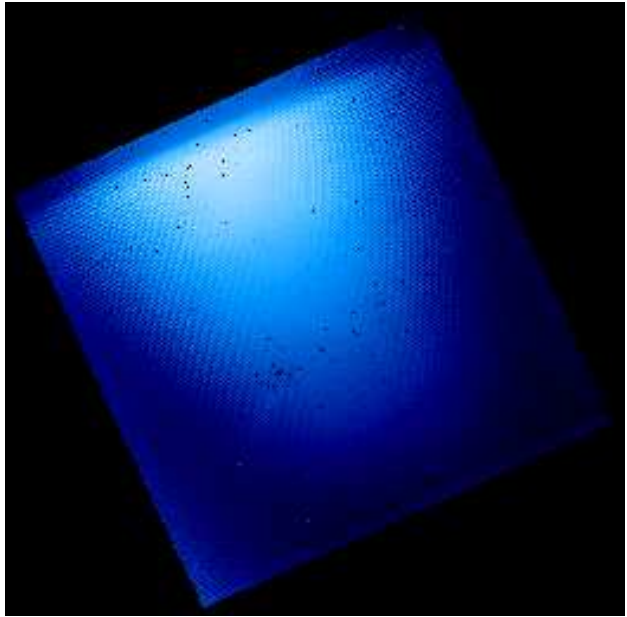}
                \caption{Box Aperture method at 2.107 $\mu$m}
                \label{fig:cube1}
\end{subfigure} \\

\begin{subfigure}[b]{0.48\textwidth}
                \includegraphics[trim = 15mm 8mm 17mm 5mm,clip,width=\textwidth]{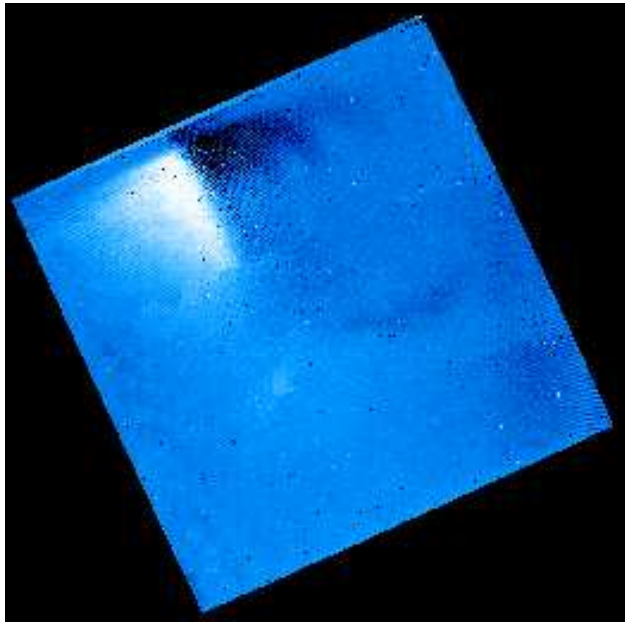}
                \caption{Microlens Inversion method at 2.107 $\mu$m}
                \label{fig:cube2}
\end{subfigure}\\

\end{tabular}
\end{center}
\caption{\label{fig:datacubes1} Final data cube slices comparing the box aperture with the microlens inversion method on the same K2 band flat field image at 2.107 $\mu$m (i.e., the blue end of the band).  Both figures are scaled to the same flux range as Figure \ref{fig:datacubes2}.  The ``checkerboard'' pattern is quite noticeable in the box aperture method. There is also significant ghosting in Figure \ref{fig:cube1} from neighboring lenslets contaminating the extraction, which is reduced to a smaller region in Figure \ref{fig:cube2} with the microlens inversion.  A real faint noise source caused by a stray light feature in the IFS is only discernible by eye through the microlens inversion technique in Figure \ref{fig:cube2} due to properly weighting the extraction. The gradient from top to bottom is due to thermal background emission, and the darker region along the top is from an internal cold stop baffle that blocks the thermal background near that edge of the detector. This is not vignetting or a low flat-field response along that edge.} 
\end{figure}

\begin{figure}
\begin{center}
\begin{tabular}{c}

\begin{subfigure}[b]{0.48\textwidth}
                \includegraphics[trim = 15mm 8mm 17mm 5mm,clip,width=\textwidth]{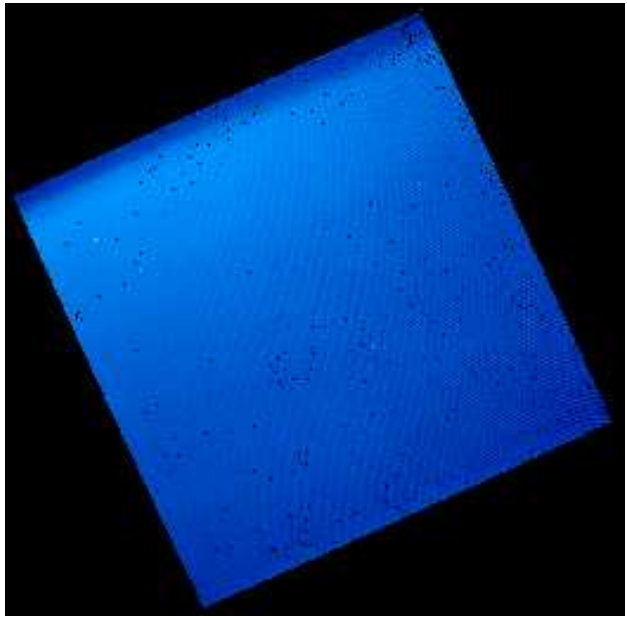}
                \caption{Box Aperture method at 2.228 $\mu$m}
                \label{fig:cube3}
\end{subfigure} \\

\begin{subfigure}[b]{0.48\textwidth}
                \includegraphics[trim = 15mm 8mm 17mm 5mm,clip,width=\textwidth]{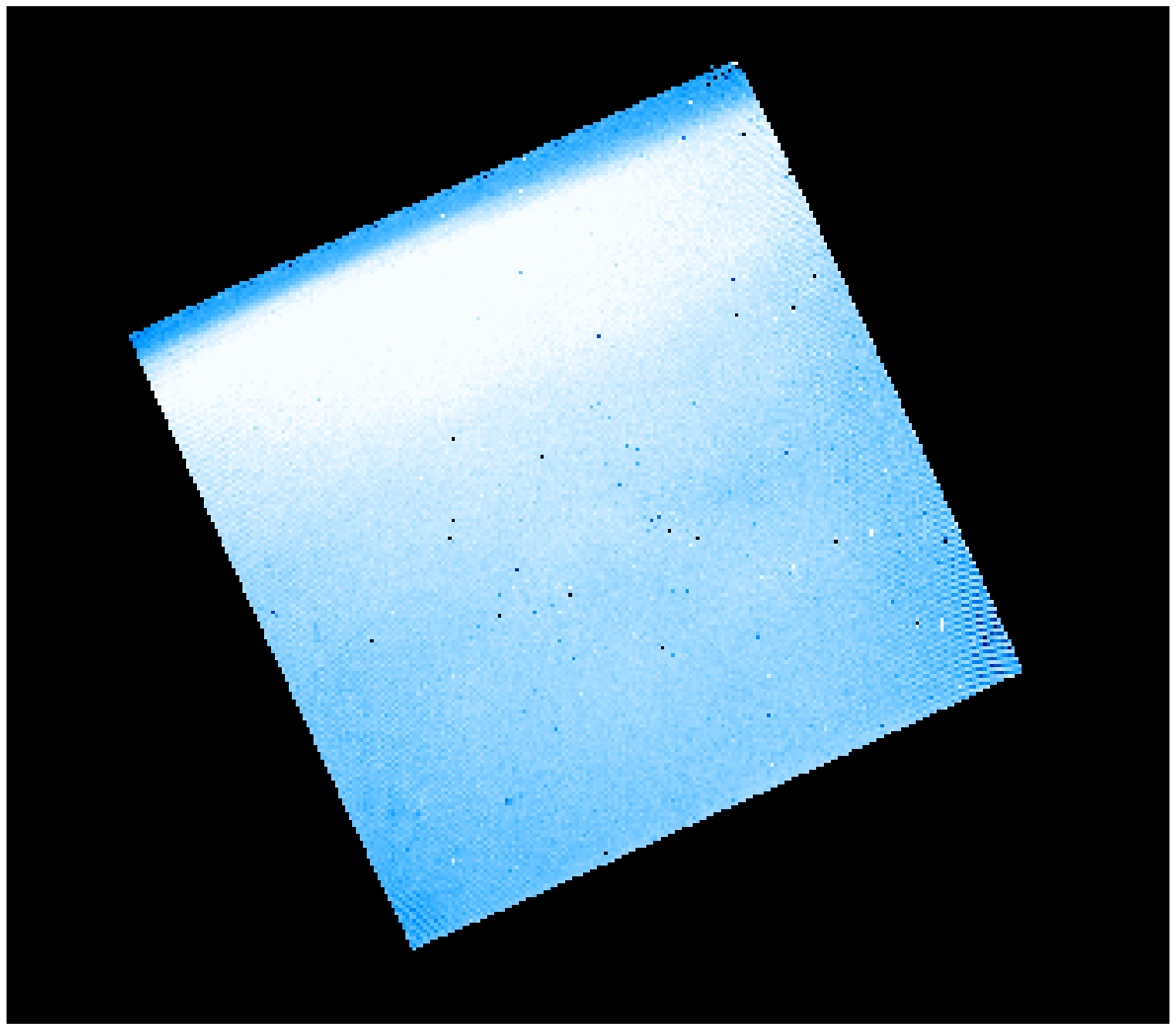}
                \caption{Microlens Inversion method at 2.228 $\mu$m}
                \label{fig:cube4}
\end{subfigure}\\

\end{tabular}
\end{center}
\caption{\label{fig:datacubes2} Final data cube slices showing the same K2 band detector flat field images as Figure \ref{fig:datacubes1}. The figures show the same wavelength slice at 2.228 $\mu$m (i.e., central region of the band) and are scaled to the same flux range between them and Figure \ref{fig:datacubes1}.  The microlens inversion slice show an increase in flux extraction, reduced bad lenslet pixelization, and reduced ``checkerboarding'' from alternating sub-pixel positions. The gradient from top to bottom is due to thermal background emission, and the darker region along the top is from an internal cold stop baffle that blocks the thermal background near that edge of the detector. This is not vignetting or a low flat-field response along that edge.}
\end{figure}

The final result of the methods outlined here can be seen in comparison to the box aperture method in Figure \ref{fig:datacubes1} and Figure \ref{fig:datacubes2} on a K2 flat field image.  Figures \ref{fig:datacubes1} and \ref{fig:datacubes2} are scaled to the same flux range and have noticeably higher flux extraction in part due to 2D cross correlation finding the flexure offset prior to extraction.  Given no initial offset, the routine found an offset of 0.77 pixels in x and -2.45 pixels in y.  Figure \ref{fig:datacubes1} illustrate how the ghosting of flux at the end of the band (2.107 $\mu$m in K2) are reduced using the inversion method to decorrelate contaminating flux from other lenslets. Figure \ref{fig:datacubes2} illustrate the use of bad pixel masking during the reduction which minimizes extraneous values in an image slice, seen as fewer dark lenslets.  Furthermore, the ``checkerboard'' aliasing pattern from alternating sub-pixel positioning for adjacent lenslets is reduced. A real caustic stray light feature in Figure \ref{fig:cube2} can be seen in the microlens inversion extraction, but not in the box aperture method.  This is likely due to the systematic noise induced by the aperture method when compared to properly weighting a PSF extraction to resolve fainter features.  When the methods outlined here have been refined, the preferred method for science quality results will use these flux extraction techniques to minimize systematic error and noise propagating from the detector to the data cubes for PSF reduction and spectral/polarimetric characterization of targets.

\acknowledgments

The GPI project has been supported by Gemini Observatory, which is operated by AURA, Inc., under a cooperative agreement with the NSF on behalf of the Gemini partnership: the NSF (USA), the National Research Council (Canada), CONICYT (Chile), the Australian Research Council (Australia), MCTI (Brazil) and MINCYT (Argentina).


\bibliography{report}   
\bibliographystyle{spiebib}   

\newpage
\clearpage

\newpage
\clearpage

\newpage
\clearpage

\end{document}